\documentclass{article}%
\usepackage{amsmath}
\usepackage{amsfonts}
\usepackage{amssymb}%
\usepackage{graphicx}
\begin{document}

\title{Comments on ``Dirac theory in spacetime algebra''}
\author{William E. Baylis\\Physics Department, University of Windsor,\\Windsor, ON, Canada N9B 3P4}
\date{April 29, 2001 (revised February 7, 2002)}
\maketitle

\begin{abstract}
In contrast to formulations of the Dirac theory by Hestenes and by the current
author, the formulation recently presented by W. P. Joyce [J. Phys. A: Math.
Gen. \textbf{34} (2001) 1991--2005] is equivalent to the usual Dirac equation
only in the case of vanishing mass. For nonzero mass, solutions to Joyce's
equation can be solutions either of the Dirac equation in the Hestenes form or
of the same equation with the sign of the mass reversed, and in general they
are mixtures of the two possibilities. Because of this relationship, Joyce
obtains twice as many linearly independent plane-wave solutions for a given
momentum eigenvalue as exist in the conventional theory. A misconception about
the symmetry of the Hestenes equation and the geometric significance of the
algebraic spinors is also briefly discussed.

\end{abstract}

The recent formulation by Joyce\cite{Joyce01} of what he calls the
\textquotedblleft bivector Dirac equation\textquotedblright\ in the complex
Dirac algebra (Clifford's geometric algebra of Minkowski spacetime taken over
the complex field) can be decomposed into the Dirac equation in the Hestenes
form\cite{Hes66,Hes75,Hes81,Hes96} for mass $+m$ plus another equation for
mass $-m$. The purpose of this comment is to prove this claim and to discuss a
few other aspects of the paper. The approach exploits the power of projectors
in geometric algebras. To start, we first review the Dirac equation for a free
electron and its formulation in algebraic form.

\section{Dirac Equation}

Joyce's formulation is made in the framework of the complex Dirac algebra,
that is, Clifford's geometric algebra\cite{Bay96,Loun97} of Minkowski
spacetime taken over the complex field. A matrix form of this algebra is
traditionally used for the Dirac equation. Since the objective is to find a
new formulation of the Dirac equation, it is reasonable first to put the
traditional formulation directly in algebraic form. For simplicity, we
consider only the case of a free electron; the procedure for adding a gauge
potential is straightforward and can be undertaken later. The Dirac equation
for a free electron in units with $\hbar=c=1$ can be written as a matrix
equation
\begin{equation}
\mathrm{i}\nabla\psi=m\psi\label{Diraceq}%
\end{equation}
where $\psi$ is a four-component complex spinor, and the gradient operator in
Minkowski spacetime (sum repeated indices $\mu=0,1,2,3$)%
\begin{equation}
\nabla=\gamma_{\mu}\partial^{\mu} \label{gradient}%
\end{equation}
(which Joyce denotes by $\mathbf{d}\,$)\footnote{Joyce has introduced new
notation for a number of terms and has redefined other symbols already in
common use. A change that is particularly liable to cause confusion is his
redefinition of the wedge and dot products. This comment employs the more
established notation of Hestenes, but I will relate this to the notation of
Joyce where appropriate.} is an expansion in the $4\times4$ gamma matrices
$\gamma_{\mu}$ with operator coefficients $\partial^{\mu}=\partial/\partial
x_{\mu}=\eta^{\mu\nu}\partial/\partial x^{\nu}=\eta^{\mu\nu}\partial_{\nu}.$
Indices are raised and lowered by the metric tensors $\left(  \eta^{\mu\nu
}\right)  $ and $\left(  \eta_{\mu\nu}\right)  $ of Minkowski spacetime. The
algebra of the gamma matrices is that of the vector basis of Clifford's
geometric algebra of spacetime, namely%
\begin{equation}
\gamma_{\mu}\gamma_{\nu}+\gamma_{\nu}\gamma_{\mu}=2\eta_{\mu\nu}\,.
\label{Clifford}%
\end{equation}
In the algebra, the orthonormal basis vectors of spacetime are therefore taken
to be $\left\{  \gamma_{0},\gamma_{1},\gamma_{2},\gamma_{3}\right\}  ,$ and
the explicit matrices that represent them are not important. Joyce has
considered two possible choices for the metric, but to avoid unnecessary
complications, I take $\left(  \eta^{\mu\nu}\right)  $ to be the diagonal
matrix%
\begin{equation}
\left(  \eta^{\mu\nu}\right)  =\operatorname*{diag}\left(  1,-1,-1,-1\right)
=\left(  \eta_{\mu\nu}\right)  \label{metric}%
\end{equation}
corresponding to the choice of Joyce's parameter $\eta=-1.$

The geometric algebra generated by associative products of basis vectors
$\gamma_{0},\gamma_{1},\gamma_{2},\gamma_{3}$ that satisfy (\ref{Clifford})
with (\ref{metric}) is denoted\cite{Loun97} $C\!\ell_{1,3}\,.$ Four spinors
$\psi,$ each satisfying the Dirac equation (\ref{Diraceq}), can be used to
construct a $4\times4$ solution matrix $\Psi$. Let each column of $\Psi$ be a
column spinor\thinspace$\psi$ that satisfies (\ref{Diraceq}). Then $\Psi$ is
the matrix representation of an element of the Dirac algebra $C\!\ell_{1,3}$
over the complex field $\mathbb{C}.$ We construct explicit plane-wave
solutions below, but the point made here is that the Dirac equation
(\ref{Diraceq}) can be written directly as an \emph{algebraic equation }in
$C\!\ell_{1,3}\otimes\mathbb{C}$
\begin{equation}
\mathrm{i}\nabla\Psi=m\Psi,\quad\Psi\in C\!\ell_{1,3}\otimes\mathbb{C}
\label{Dirac}%
\end{equation}
that gives up to four independent solutions of the traditional column-spinor
equation. This equation remains valid under multiplication from right by any
constant algebraic element. In particular, one can multiply by a projector
$\mathsf{P}$, that is a hermitian idempotent%
\begin{equation}
\mathsf{P}^{2}=\mathsf{P}=\mathsf{P}^{\dag}.
\end{equation}
For example, columns of a $4\times4$ matrix representation of the algebraic
spinor $\Psi$ can be extracted by the application of projectors $\mathsf{P}%
\left(  \alpha\right)  ,\,\alpha=0,1,2,3$ defined with diagonal matrix
representations $\mathsf{P}\left(  \alpha\right)  =\left(  \delta_{\mu\alpha
}\delta_{\nu\alpha}\right)  \,.$ Thus, the algebraic equation (\ref{Dirac}) is
equivalent to four copies of the usual spinor equation (\ref{Diraceq}).

The projectors $\mathsf{P}\left(  \alpha\right)  $ depend on the matrix
representation. However, as we see below, useful projectors can also be
defined algebraically, independent of the representation.

\section{Joyce Equation}

Joyce has formulated a \emph{different} equation in the complex Dirac algebra
$C\!\ell_{1,3}\otimes\mathbb{C}.$ His Eq.(19) for a free electron is%
\begin{equation}
\mathrm{i}\nabla\Psi_{\mathrm{J}}=m\Psi_{\mathrm{J}}\gamma_{0}\,\quad\quad
\Psi_{\mathrm{J}}\in C\!\ell_{1,3}^{+}\otimes\mathbb{C}\, \label{JoyceDE}%
\end{equation}
although Joyce uses the notation $\mathbf{e}_{\mu}$ in place of $\gamma_{\mu
}\,.$ It differs from the Dirac equation (\ref{Dirac}) by the extra factor of
$\gamma_{0}$ on the right. The equations are the same only if $m=0.$ The Joyce
equation (\ref{JoyceDE}) is invariant under right multiplication of
$\Psi_{\mathrm{J}}$ by any constant element that commutes with $\gamma_{0}.$
The spinor can be decomposed into pieces $\Psi_{\mathrm{J}}=\Psi_{\mathrm{J}%
}\mathsf{P}_{+0}+\Psi_{\mathrm{J}}\mathsf{P}_{-0}$ where the constant elements%
\begin{equation}
\mathsf{P}_{\pm0}=\frac{1}{2}\left(  1\pm\gamma_{0}\right)
\end{equation}
are projectors. Equation (\ref{JoyceDE}) then splits into two parts, one
equivalent to the usual Dirac equation (\ref{Diraceq}) in algebraic form, and
the other similar but with the sign of the mass changed:%
\begin{equation}
\mathrm{i}\nabla\Psi_{\mathrm{J}}\mathsf{P}_{\pm0}=\pm m\Psi_{\mathrm{J}%
}\mathsf{P}_{\pm0}%
\end{equation}
where we noted the property that $\gamma_{0}\mathsf{P}_{\pm0}=\pm
\mathsf{P}_{\pm0}\,.$ Thus, the part $\Psi_{\mathrm{J}}\mathsf{P}_{+0}$ is a
solution to the Dirac equation (\ref{Dirac}), but $\Psi_{\mathrm{J}}%
\mathsf{P}_{-0}$ satisfies an equation with the opposite mass $-m.$ However as
indicated in equation (\ref{JoyceDE}), Joyce has imposed an additional
condition on his solutions $\Psi_{\mathrm{J}}$, namely that they be elements
of the \emph{even} subalgebra $C\!\ell_{1,3}^{+}\otimes\mathbb{C}$ of the
complex Dirac algebra (he refers to $\Psi_{\mathrm{J}}$ as a \textquotedblleft
generalized bivector\textquotedblright, although it can also contain scalars
and 4-volume elements). Thus $\Psi_{\mathrm{J}}$ can be expanded%
\begin{equation}
\Psi_{\mathrm{J}}=\Psi_{\mathrm{J}}^{K}\,\Gamma_{K}^{+}\,,\;K=1,2,\ldots,8
\label{expansn}%
\end{equation}
in the basis $\left\{  \Gamma_{K}^{+}\right\}  =\left\{  1,\gamma_{\mu}%
\gamma_{\nu},\gamma_{0}\gamma_{1}\gamma_{2}\gamma_{3}\right\}  $ of
$C\!\ell_{1,3}^{+}$ with $\mu,\nu=0,1,2,3$ and $\mu<\nu,$ where $\Psi
_{\mathrm{J}}^{K}$ are complex scalar functions. However, the part
$\Psi_{\mathrm{J}}\mathsf{P}_{+0}$ of the Joyce spinor $\Psi_{\mathrm{J}}$
that is a solution of the Dirac equation (\ref{Dirac}) cannot be even since if
$\Psi_{\mathrm{J}}$ is even, then $\Psi_{\mathrm{J}}\gamma_{0}$, which is part
of $\Psi_{\mathrm{J}}\mathsf{P}_{+0}$, is odd. Indeed, the even and odd parts
of $\Psi_{\mathrm{J}}\mathsf{P}_{+0}$ are of the same size. Consequently,
there is \emph{no} solution $\Psi_{\mathrm{J}}\in$ $C\!\ell_{1,3}^{+}%
\otimes\mathbb{C}$ that is also a solution of the algebraic Dirac equation
(\ref{Dirac}) if $m>0\,.$ Nevertheless, this result does not preclude the
possibility that there is some other relation between $\Psi_{\mathrm{J}}$ and
$\psi$ that would be consistent with both equations (\ref{Diraceq}) and
(\ref{JoyceDE}). Joyce does not specify an explicit relation between his
$\Psi_{\mathrm{J}}$ and the usual Dirac spinor, but we can find one by
comparing his equation (\ref{JoyceDE}) with that of the Hestenes form.

\section{Hestenes Equation}

Hestenes\cite{Hes66,Hes96} constructed a form of the Dirac equation in the
\emph{real} algebra $C\!\ell_{1,3},$ which he calls the spacetime algebra
(STA). The Hestenes equation for the free electron in the STA is%
\begin{equation}
-\nabla\Psi_{\mathrm{H}}\gamma_{1}\gamma_{2}=m\Psi_{\mathrm{H}}\gamma
_{0}\,,\quad\Psi_{\mathrm{H}}\in C\!\ell_{1,3}^{+}\,. \label{HestenesDE}%
\end{equation}
The Joyce and Hestenes equations act in different spaces and are not
equivalent. To compare them, we consider both acting in the larger space of
the complex Dirac algebra $C\!\ell_{1,3}\otimes\mathbb{C~}$.

The Joyce and Hestenes spinors in $C\!\ell_{1,3}\otimes\mathbb{C}$ are related
by the simple projectors%
\begin{equation}
\mathsf{P}_{\pm12}=\frac{1}{2}\left(  1\pm\mathrm{i}\gamma_{1}\gamma
_{2}\right)  \label{P12}%
\end{equation}
which can be applied to the Joyce equation (\ref{JoyceDE}) to give%
\begin{equation}
\mathrm{i}\nabla\left(  \Psi_{\mathrm{J}}\mathsf{P}_{\pm12}\right)  =\mp
\nabla\left(  \Psi_{\mathrm{J}}\mathsf{P}_{\pm12}\right)  \gamma_{1}\gamma
_{2}=m\left(  \Psi_{\mathrm{J}}\mathsf{P}_{\pm12}\right)  \gamma_{0}%
\end{equation}
where we noted%
\begin{equation}
\mathsf{P}_{\pm12}=\pm\mathrm{i}\mathsf{P}_{\pm12}\gamma_{1}\gamma_{2}\,.
\label{Pacwoman}%
\end{equation}
Thus, any spinor solution $\Psi_{\mathrm{J}}$ can be split into parts%
\begin{equation}
\Psi_{\mathrm{J}}=\Psi_{\mathrm{J}}\mathsf{P}_{+12}+\Psi_{\mathrm{J}%
}\mathsf{P}_{-12}%
\end{equation}
one of which ($\Psi_{\mathrm{J}}\mathsf{P}_{+12}$) satisfies the Dirac
equation in the Hestenes form and the other of which ($\Psi_{\mathrm{J}%
}\mathsf{P}_{-12}$) satisfies the same equation but with the sign of the mass reversed.

As noted above, $\Psi_{\mathrm{J}}$ is generally complex whereas the Hestenes
$\Psi_{\mathrm{H}}$ is real. Let $\Psi_{\mathrm{J}}^{\ast}$ be the complex
conjugate of $\Psi_{\mathrm{J}},$ obtained by replacing each scalar
coefficient $\Psi_{\mathrm{J}}^{K}$ in the expansion (\ref{expansn}) of
$\Psi_{\mathrm{J}}$ by its complex conjugate. The property (\ref{Pacwoman})
can be used to replace the complex even $\Psi_{\mathrm{J}}$ in the products
$\Psi_{\mathrm{J}}\mathsf{P}_{\pm12}$ by real even elements $\Psi_{\mathrm{J}%
}^{\pm}~$:%
\begin{align}
\Psi_{\mathrm{J}}\mathsf{P}_{\pm12}  &  =\Psi_{\mathrm{J}}^{\pm}%
\mathsf{P}_{\pm12}\\
\Psi_{\mathrm{J}}^{\pm}  &  \equiv\frac{1}{2}\left(  \Psi_{\mathrm{J}}%
+\Psi_{\mathrm{J}}^{\ast}\right)  \pm\frac{\mathrm{i}}{2}\left(
\Psi_{\mathrm{J}}-\Psi_{\mathrm{J}}^{\ast}\right)  \gamma_{1}\gamma_{2}
\label{PsiJpm}%
\end{align}
thereby giving $\Psi_{\mathrm{J}}^{\pm}\mathsf{P}_{\pm12}$ as solutions of the
Hestenes form of the Dirac equation with the correct or opposite sign on the
mass term. Note that $\mathsf{P}_{\pm12}$ is even and therefore preserves the
even property of the spinor. It is also complex, but since the Hestenes
equation (\ref{HestenesDE}) is real and linear, the real part of any complex
solution is also a solution. The real part of $\Psi_{\mathrm{J}}^{\pm
}\mathsf{P}_{\pm12}$ is just $\frac{1}{2}\Psi_{\mathrm{J}}^{\pm}$, and
consequently the two spinors $\Psi_{\mathrm{J}}^{\pm}$ (\ref{PsiJpm}) are real
even solutions of the Dirac equation in the Hestenes form (\ref{HestenesDE})
with the two signs of the mass term. The products $\Psi_{\mathrm{J}}^{\pm
}\gamma_{1}\gamma_{2}$ are similar solutions. The sum and difference of such
solutions give the real and imaginary parts of the Joyce spinor $\Psi
_{\mathrm{J}}~$:%
\begin{equation}
\Psi_{\mathrm{J}}=\frac{1}{2}\left(  \Psi_{\mathrm{J}}^{+}+\Psi_{\mathrm{J}%
}^{-}\right)  +\frac{\mathrm{i}}{2}\left(  \Psi_{\mathrm{J}}^{+}%
-\Psi_{\mathrm{J}}^{-}\right)  \gamma_{1}\gamma_{2}~.
\end{equation}
This completes the demonstration relating every solution of the Joyce equation
(\ref{JoyceDE}) to sums and differences of solutions to the Hestenes form of
the Dirac equation with correct and reversed signs on the mass term. The close
association of Hestenes and Joyce spinors is reasonable considering that the
current density $\mathbf{J}$ has the same form, namely the vector part of
$\Psi\gamma_{0}\widetilde{\Psi}$ ($\widetilde{\Psi}$ is the reversion of
$\Psi$), if $\Psi$ belongs to the real algebra $C\!\ell_{1,3}\,.$

It is noted in passing that similar arguments, using both pairs of commuting
projectors $\mathsf{P}_{\pm12}$ and $\mathsf{P}_{\pm0}~$, can relate a general
solution $\Psi$ of the Dirac equation (\ref{Dirac}) to four independent
solutions $\Psi_{\mathrm{H}}\in C\!\ell_{1,3}^{+}$ of the Hestenes equation
(\ref{HestenesDE}) (with the correct mass term):%
\begin{align}
\Psi_{\mathrm{H1}}  &  =\frac{1}{2}\left(  \Psi_{+}+\Psi_{+}^{\ast}\right)
\gamma_{1}\gamma_{2}+\frac{1}{2\mathrm{i}}\left(  \Psi_{-}-\Psi_{-}^{\ast
}\right)  \gamma_{0}~,\ \Psi_{\mathrm{H2}}=\Psi_{\mathrm{H1}}\gamma_{1}%
\gamma_{2}\nonumber\\
\Psi_{\mathrm{H3}}  &  =\frac{\mathrm{i}}{2}\left(  \Psi_{+}-\Psi_{+}^{\ast
}\right)  \gamma_{1}\gamma_{2}+\frac{1}{2}\left(  \Psi_{-}+\Psi_{-}^{\ast
}\right)  \gamma_{0}~,\ \Psi_{\mathrm{H4}}=\Psi_{\mathrm{H3}}\gamma_{1}%
\gamma_{2}%
\end{align}
where $\Psi_{+}$ and $\Psi_{-}$ are the even and odd parts of $\Psi.$

\section{Symmetry of the Hestenes Equation}

Part of Joyce's stated reason for seeking an alternative algebraic form of the
Dirac equation was that he viewed the Hestenes form (\ref{HestenesDE}) as
giving special status to given directions in space. In particular, because
equation (\ref{HestenesDE}) contains the $\gamma_{1}\gamma_{2}$ bivector, it
was felt that the corresponding plane was singled out. On this basis it might
be disappointing that every solution to the Joyce equation (\ref{JoyceDE}) is
a combination of solutions to two equations of the Hestenes form. However, the
asymmetry that Joyce saw in the Hestenes equation is only apparent, as
explained below.

The principal advantages of the Hestenes formulation of the Dirac equation are
(1) that it acts in the real Dirac algebra $C\!\ell_{1,3}$ rather than in the
more traditional complex Dirac algebra used by most authors as well as by
Joyce, and (2) it offers unambiguous geometrical interpretations for
expressions in the theory. The fact that the spinor of the Hestenes
formulation is an even element and that the Hestenes equation preserves its
evenness suggests that the Dirac theory can also be formulated in the real
Pauli algebra $C\!\ell_{3},$ which is isomorphic to $C\!\ell_{1,3}^{+}\,.$
Indeed there is a very simple covariant formulation\cite{Bay92,Bay97,Dav98}
using paravectors\cite{Bay99} of $C\!\ell_{3},$ and this is closely related to
formulations in biquaternions and $2\times2$ matrices.\cite{Lan29,Gur56,Ras88}
Further background and references can be found in a couple of recent
papers.\cite{Gsp01,Fau01}

In both the $C\!\ell_{3}$ formulation and in Hestenes' analysis, the spinor
plays the role of a relativistic transformation amplitude from the reference
frame of the fermion to the lab frame. The orientation of the reference frame
is not significant since global gauge transformations $\Psi_{\mathrm{H}%
}\rightarrow\Psi_{\mathrm{H}}R$, where $R$ is a fixed spatial rotor, can
rotate it arbitrarily. It is therefore of no physical consequence that the
particular bivector $\gamma_{1}\gamma_{2}$ appears in the Hestenes equation
(\ref{HestenesDE}).\cite{Dor96}

\section{Plane-Wave Solutions}

In Section 10 of his paper, Joyce sought plane-wave solutions of the form
$\Psi_{\mathrm{J}}=A\exp\left(  \mathrm{i}k^{\mu}x_{\mu}\right)  =A\exp\left(
\mathrm{i}\omega t-\mathrm{i}kx\right)  ,$ where $A$ is a constant element, to
his equation (\ref{JoyceDE}). Differentiation gives the
condition\footnote{Joyce's related equation (22) differs in the sign of $k$,
probably stemming from the identification of the component $k$ or the
definition of $\nabla$ (his $\mathbf{d}$), equation (2). My explicit
identification is $k\equiv k^{1}$ and $x\equiv x^{1}.$}%
\begin{equation}
-\left(  \omega+\gamma_{0}\gamma_{1}k\right)  A=m\gamma_{0}A\gamma_{0}\,.
\label{PWcondition}%
\end{equation}
Solving for $A$, substituting into the left-hand side of (\ref{PWcondition}),
and noting $\left(  \gamma_{0}\gamma_{1}\right)  ^{2}=1,$ one finds
$\omega^{2}=k^{2}+m^{2}$ for nonvanishing $A\,.$ Furthermore, $A$ has an
expansion analogous to that of $\Psi_{\mathrm{J}}\,$(\ref{expansn}) and is
conveniently split into one part that commutes with $\gamma_{0}$ and another
that anticommutes with it:%
\begin{align}
A  &  =A_{+}+A_{-}\\
\gamma_{0}A_{\pm}\gamma_{0}  &  =\pm A_{\pm}\\
A_{+}  &  =a+b\gamma_{1}\gamma_{2}+c\gamma_{2}\gamma_{3}+d\gamma_{3}\gamma
_{1}\\
A_{-}  &  =\gamma_{0}\gamma_{1}\left(  a^{\prime}+b^{\prime}\gamma_{1}%
\gamma_{2}+c^{\prime}\gamma_{2}\gamma_{3}+d^{\prime}\gamma_{3}\gamma
_{1}\right)  \,.
\end{align}
A necessary and sufficient relation for the nontrivial solution of
(\ref{PWcondition}) is easily seen to be%
\begin{equation}
A_{-}=-\frac{\omega+m}{k}\gamma_{0}\gamma_{1}A_{+}%
\end{equation}
and the equivalent relation%
\begin{equation}
A_{+}=-\frac{\omega-m}{k}\gamma_{0}\gamma_{1}A_{-}~.
\end{equation}
Except for the sign of $k$, the solution is equivalent to that found by Joyce,
and as he points out, there are eight independent solutions for a given
eigenvalue of the momentum $\mathbf{p}=k\mathbf{e}_{1}$ (four for each sign of
$\omega=\pm\sqrt{k^{2}+m^{2}}$), and this is twice as many as for the Dirac
equation. The presence of eight such solutions reflects the fact, discussed
above, that Joyce's spinor can be split into parts that solve both the usual
Dirac equation and same equation with a reversed mass sign. Explicitly, if
$b=\mathrm{i}a$ and $d=\mathrm{i}c,$ the part $\Psi_{\mathrm{J}}%
\mathsf{P}_{-12}$ of the plane-wave solution $\Psi_{\mathrm{J}}$ vanishes and
$\Psi_{\mathrm{J}}$ satisfies the usual Dirac equation, whereas if
$b=-\mathrm{i}a$ and $d=-\mathrm{i}c,$ the part $\Psi_{\mathrm{J}}%
\mathsf{P}_{+12}$ vanishes and $\Psi_{\mathrm{J}}$ satisfies the Dirac
equation with mass $-m~.$ Otherwise, the plane wave $\Psi_{\mathrm{J}}$ is a
solution of the Dirac equation only in the limit $m\rightarrow0.$

\subsection*{Acknowledgement}

The author is grateful for the hospitality of A Lasenby and the Cavendish
Astrophysics Group during a sabbatical leave spent there. He also acknowledges
helpful discussions with C Doran and the financial support of the Natural
Sciences and Engineering Research Council of Canada.

\end{document}